\documentclass[11pt,twoside]{article}

\usepackage{asp2006}
\usepackage{epsf}
\usepackage{psfig}
\usepackage{lscape}
\usepackage{epsfig, subfigure}

\begin{document}
\title{Software Data-Processing Pipeline for Transient Detection}
\author{Jayanth Chennamangalam$^1$, Yogesh Maan$^{1,2}$ and\\Avinash A. Deshpande$^1$}
\affil{$^1$Raman Research Institute, Bangalore 560 080, India  \\
$^2$Joint Astronomy Programme, Indian Institute of Science,\\Bangalore 560 012, India}

\begin{abstract}
Although several existing and upcoming telescopes have imaging as their primary mode, they also have a sensitive phased-array mode with a multiple-beam forming capability enabling high time resolution studies of several types of objects, including pulsars. For example, the potentially wide coverage in frequency, combined with its collecting area, makes the MWA-LFD a unique instrument for low-frequency detection and studies of pulsars and transients. A software data-processing pipeline is being developed by the Raman Research Institute for this purpose. We describe the various issues relevant to the detection strategies, illustrated with real data at low radio frequencies.

\end{abstract}

\section{Introduction}
\begin{figure}
 \centering
 \epsfig{figure=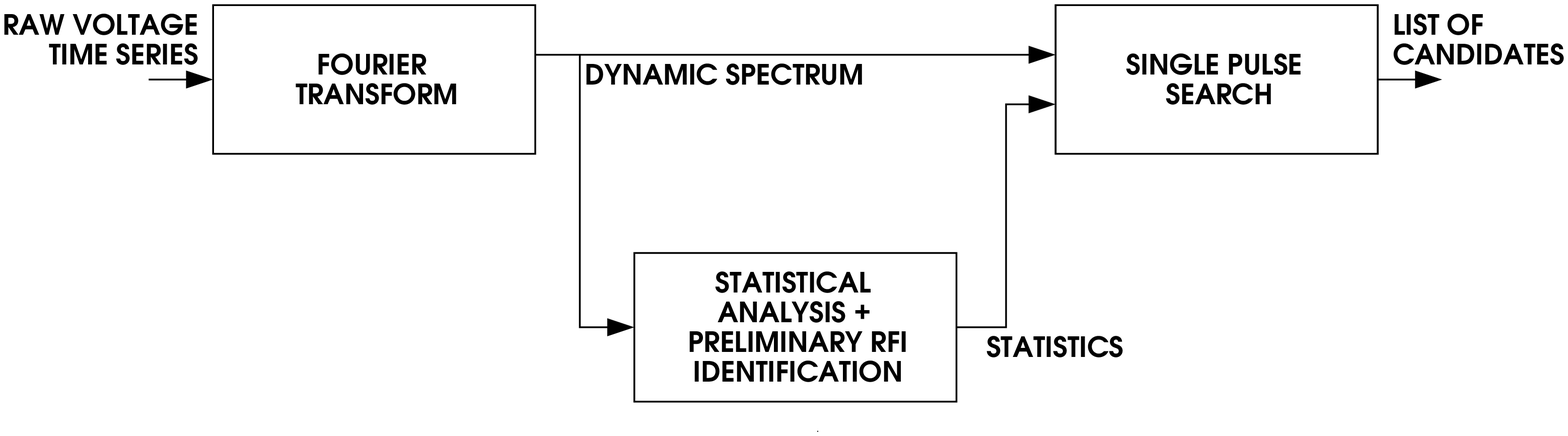,width=\linewidth}
 \caption{Schematic of the data analysis pipeline.}
 \label{fig_pipeline}
 \end{figure}

The transient radio universe has not yet been monitored and studied in detail. Upcoming telescopes such as the Murchison Widefied Array-Low Frequency Demonstrator (MWA-LFD) have studies of radio transients as one of their major science goals. A software data-processing pipeline is being developed by the Raman Research Institute for low-frequency detection and studies of transients and pulsars. A schematic of the transient search pipeline is give in Fig.~\ref{fig_pipeline}.

\section{De-dispersion}
The objective of transient search is the detection of dispersed pulses in noisy data. Therefore, the first step in the search for transients and pulsars is de-dispersion, wherein frequency-dependent delay correction is employed, and the delay-corrected frequency channels are collapsed, to achieve maximum signal to noise ratio (S/N). De-dispersion is done over a range of trial Dispersion Measure (DM) values. The delay between two frequency channels ${f_{1}}$ and ${f_{2}}$ is given, in units of milliseconds, by
\begin{equation}
\Delta t \approx 4.148808 \times 10^6 \times \left({\frac{1}{{f_1}^{2}}} - {\frac{1}{{f_2}^{2}}}\right) \times DM
\end{equation}
where the frequencies are in MHz and DM is expressed in cm${^{-3}}$ pc. The DM value is stepped over
a range of trial DMs, with the step size corresponding to minimum smearing of the pulse given by
\begin{equation}
\Delta DM = 1.205 \times 10^{-7} \times \left({\frac{{f_c}^3}{\Delta f}}\right) \times t_s
\end{equation}
where ${f_c}$ is the centre frequency of the band and ${\Delta f}$ is the bandwidth, both in MHz, and ${t_s}$ is the sampling period in milliseconds.

\section{Single Pulse Search}
The single pulse search technique is a brute-force search for S/N as a function of DM and smoothing. Unlike the standard pulsar search technique, periodicity is not probed in this method. The process involves searching for single bright pulses \citep{cordes:2003:apj, maan:2007:asixxv}.

\subsection{Threshold Detection}
If the intensity of a sample crosses a particular threshold that is not crossed by its noise component alone, it is a potential candidate. Selection of the threshold is an important criterion in transient search. If the threshold is too low, inherent noise fluctuation can cause a large number of false alarms. If the threshold is too high, a relatively weak astronomical signal may not be registered as a candidate. For Gaussian noise, the threshold ${\eta}$ in units of RMS noise is given by
\begin{equation}
erf \frac{\eta}{\sqrt{2}} = 1 - 2 \times \frac{N_{fa}}{N}
\end{equation}
where ${N_{fa}}$ is the number of false alarms and ${N}$ is the total number of samples, and ${erf}$ is the
error function whose standard form is given by
\begin{equation}
erf \ u = \frac{2}{\sqrt{\pi}} \times \int_0^u e^{-u^2} du
\end{equation}

\section{Radio Frequency Interference}
Radio frequency interference (RFI) can have a significant impact on the ability to detect weak astronomical signals. RFI is mainly classified based on their bandwidth.

 \begin{figure}
 \centering
 \subfigure[]{
 \epsfig{figure=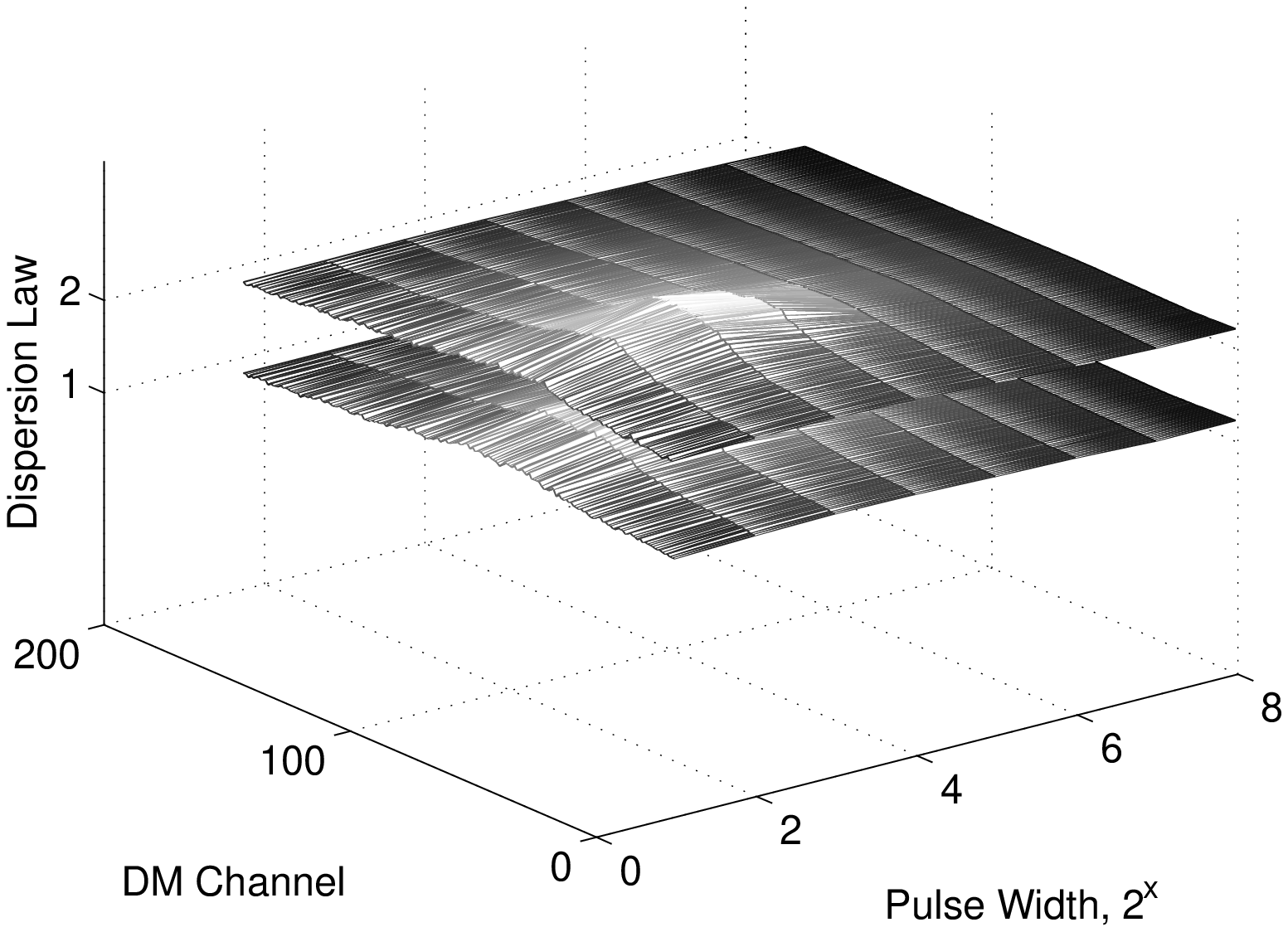,width=0.4\linewidth}}
 \subfigure[]{
 \epsfig{figure=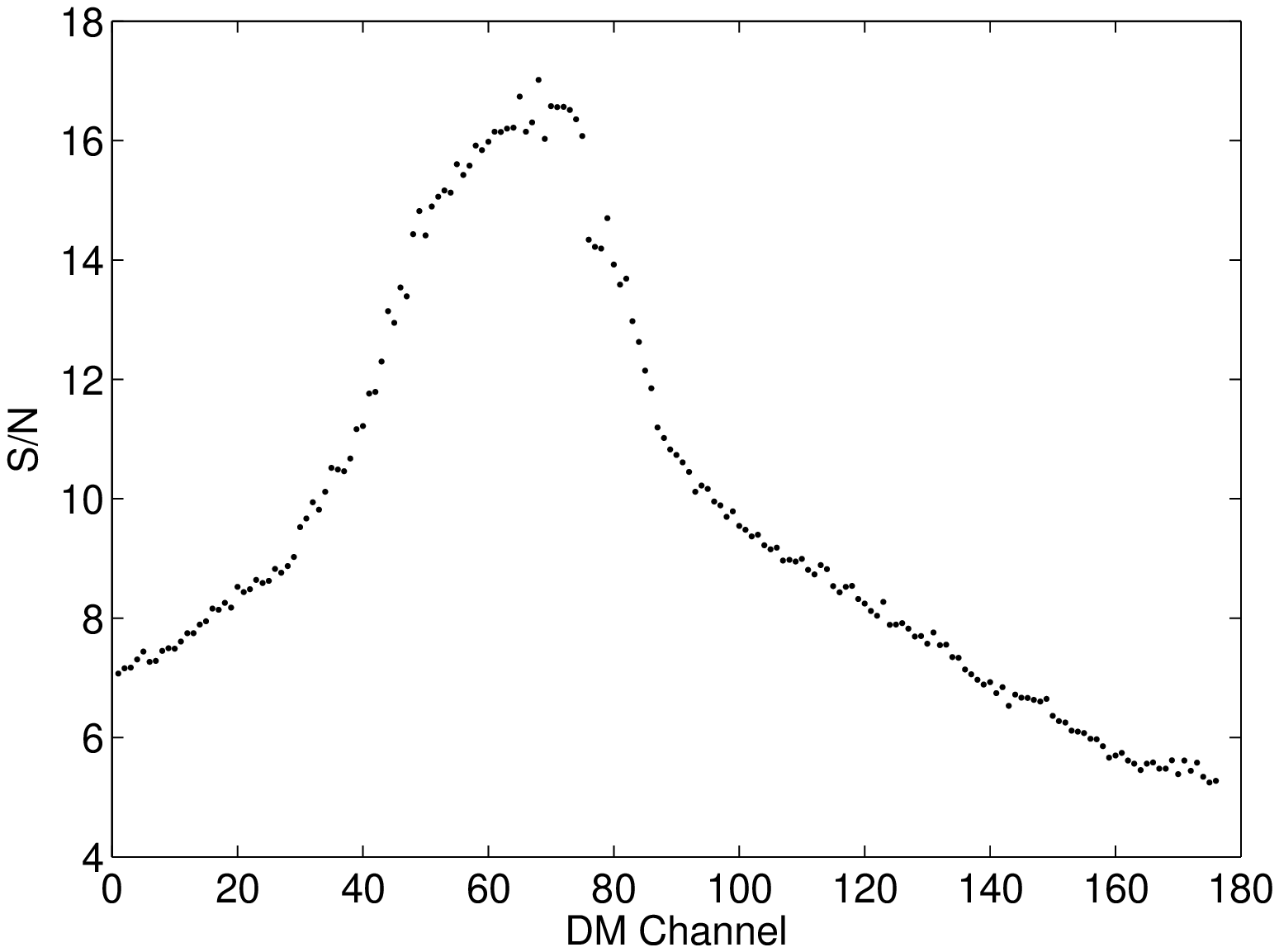,width=0.4\linewidth}}
 \subfigure[]{
 \epsfig{figure=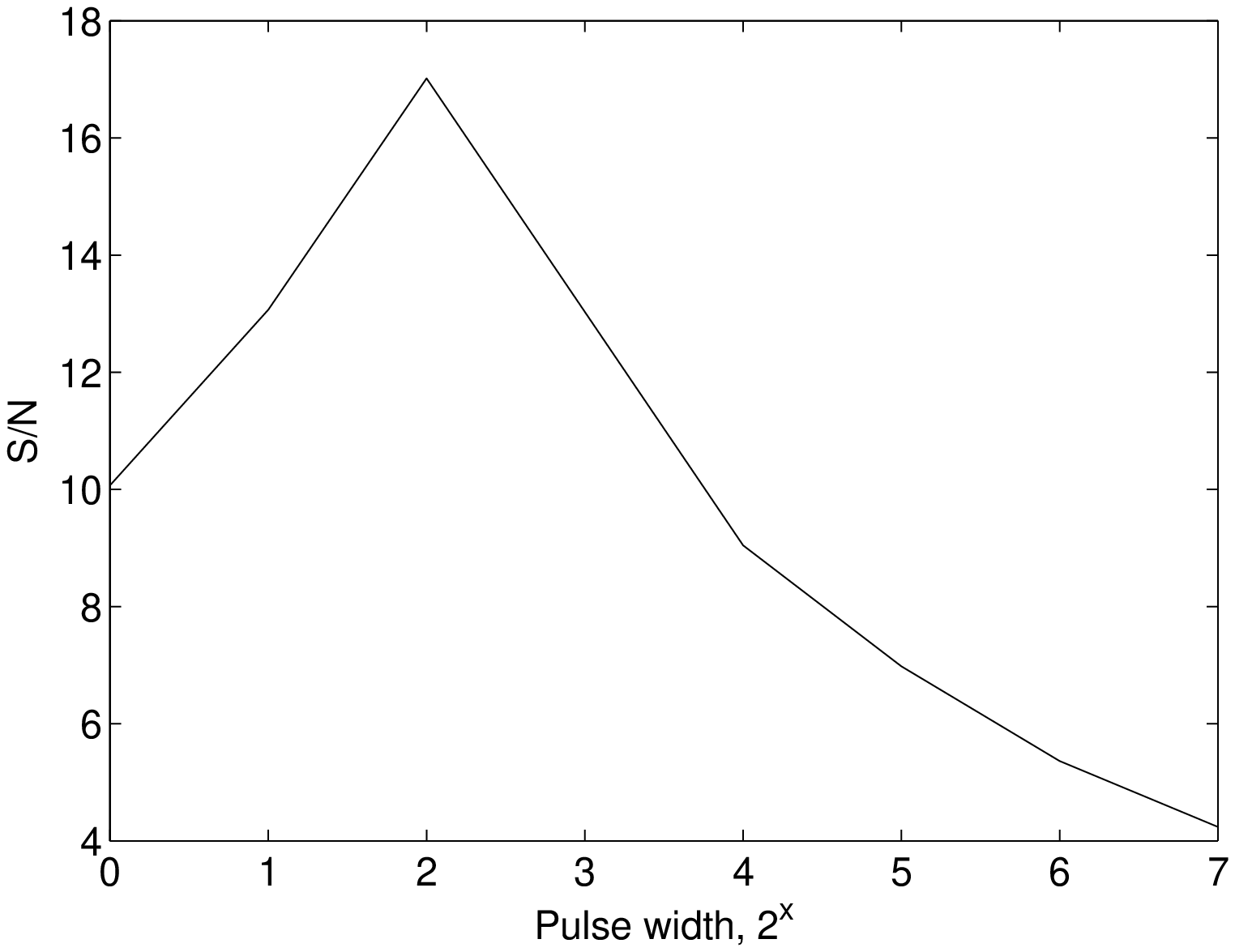,width=0.4\linewidth}}
 \subfigure[]{
 \epsfig{figure=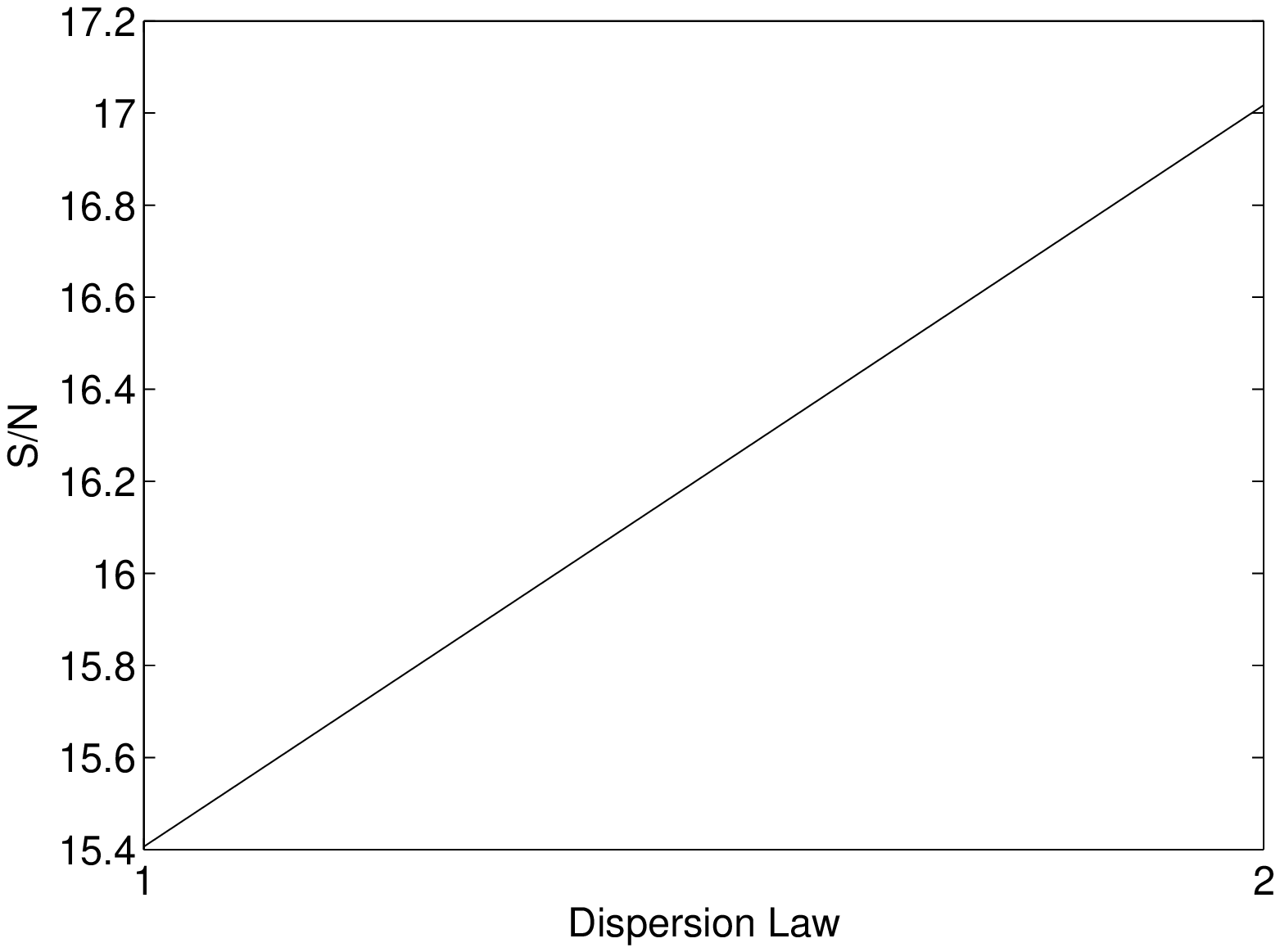,width=0.4\linewidth}}
 \subfigure[]{
 \epsfig{figure=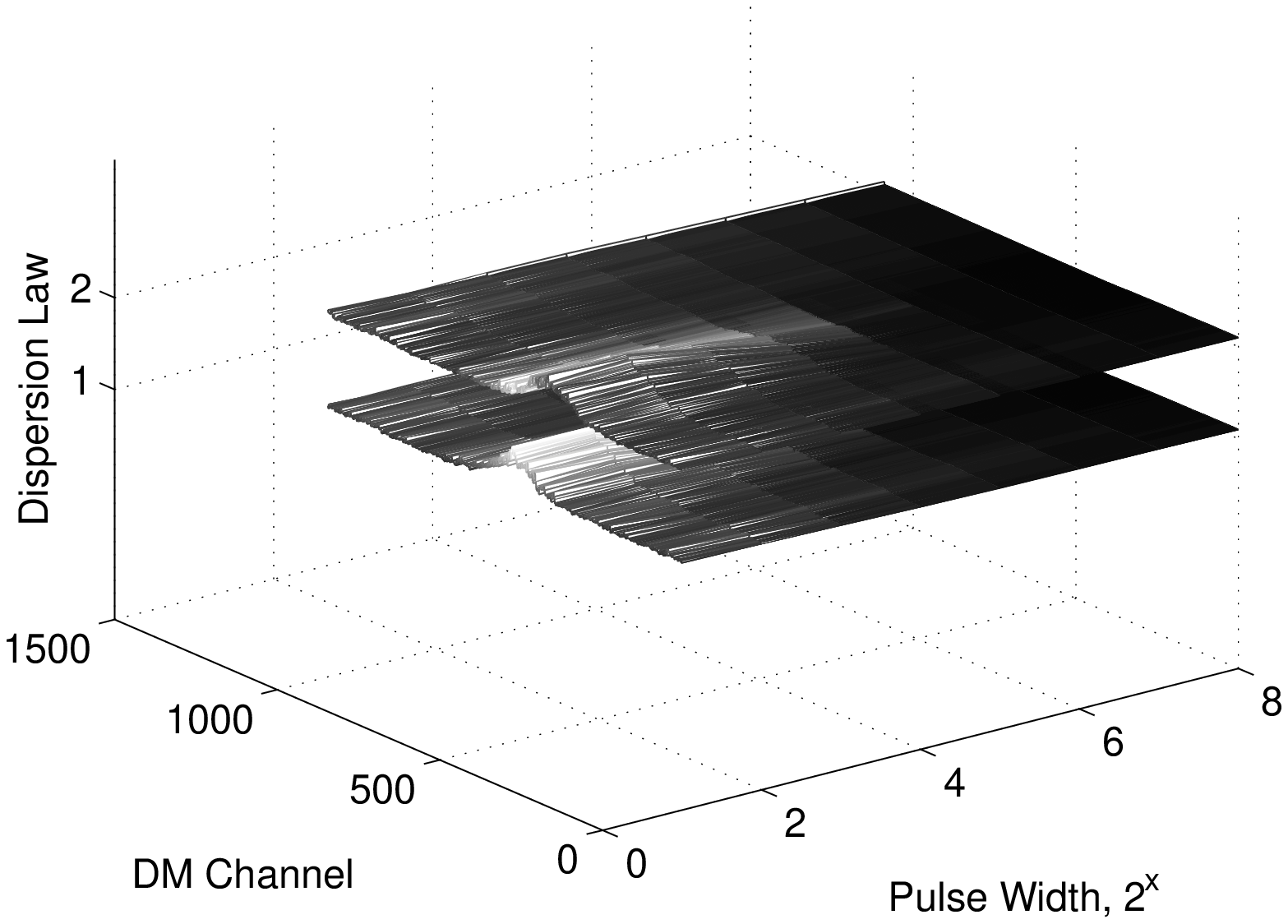,width=0.4\linewidth}}
 \subfigure[]{
 \epsfig{figure=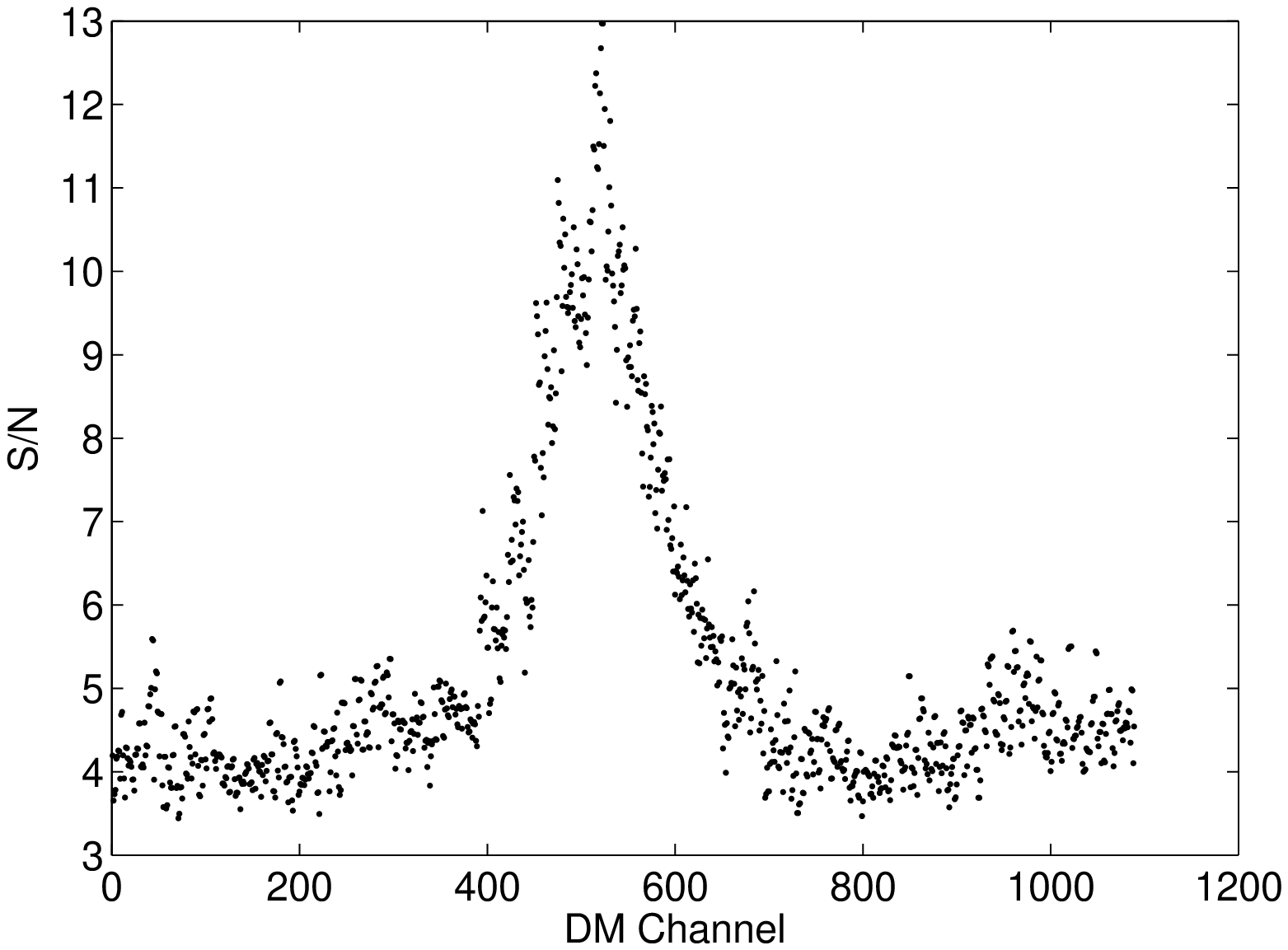,width=0.4\linewidth}}
 \subfigure[]{
 \epsfig{figure=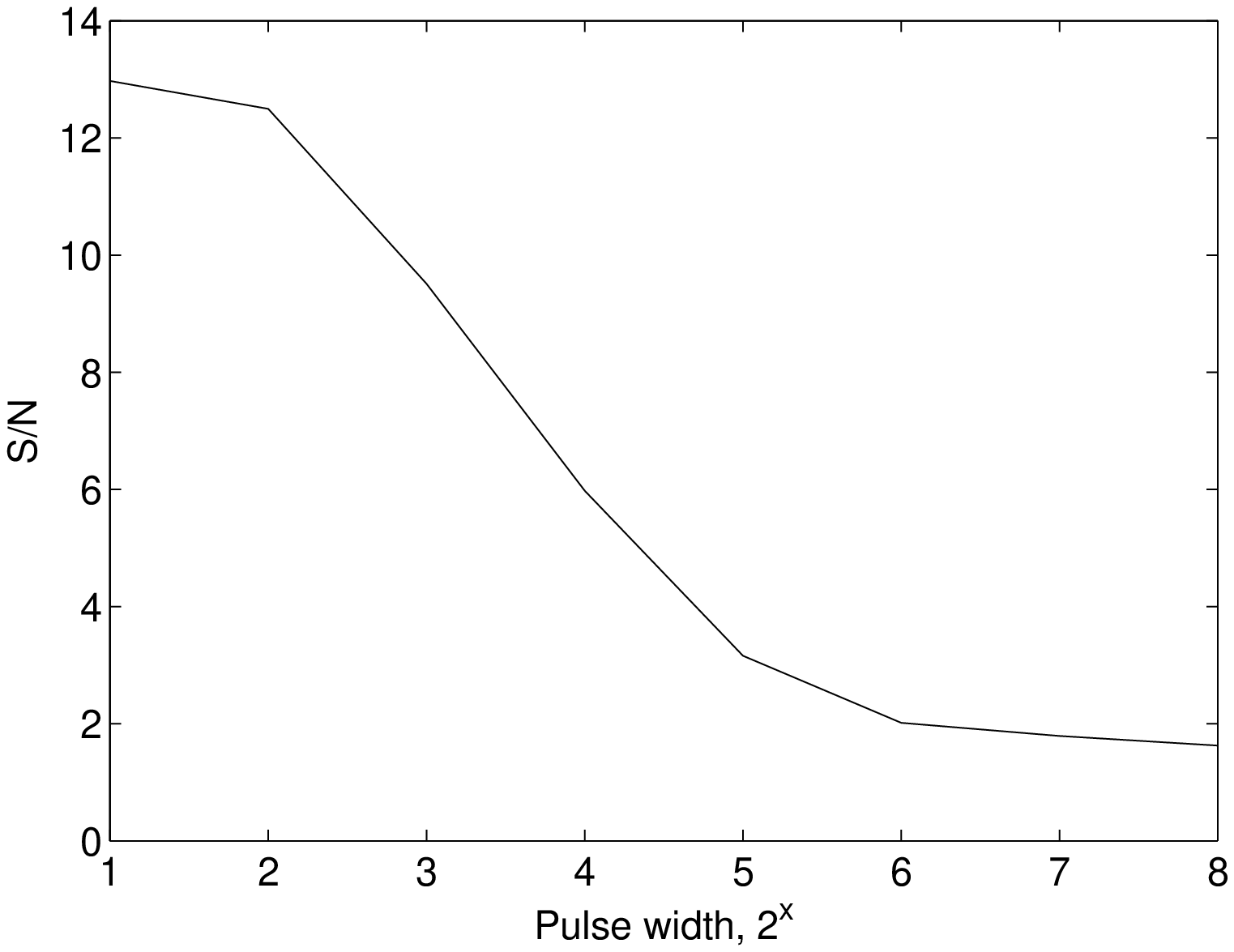,width=0.4\linewidth}}
 \subfigure[]{
 \epsfig{figure=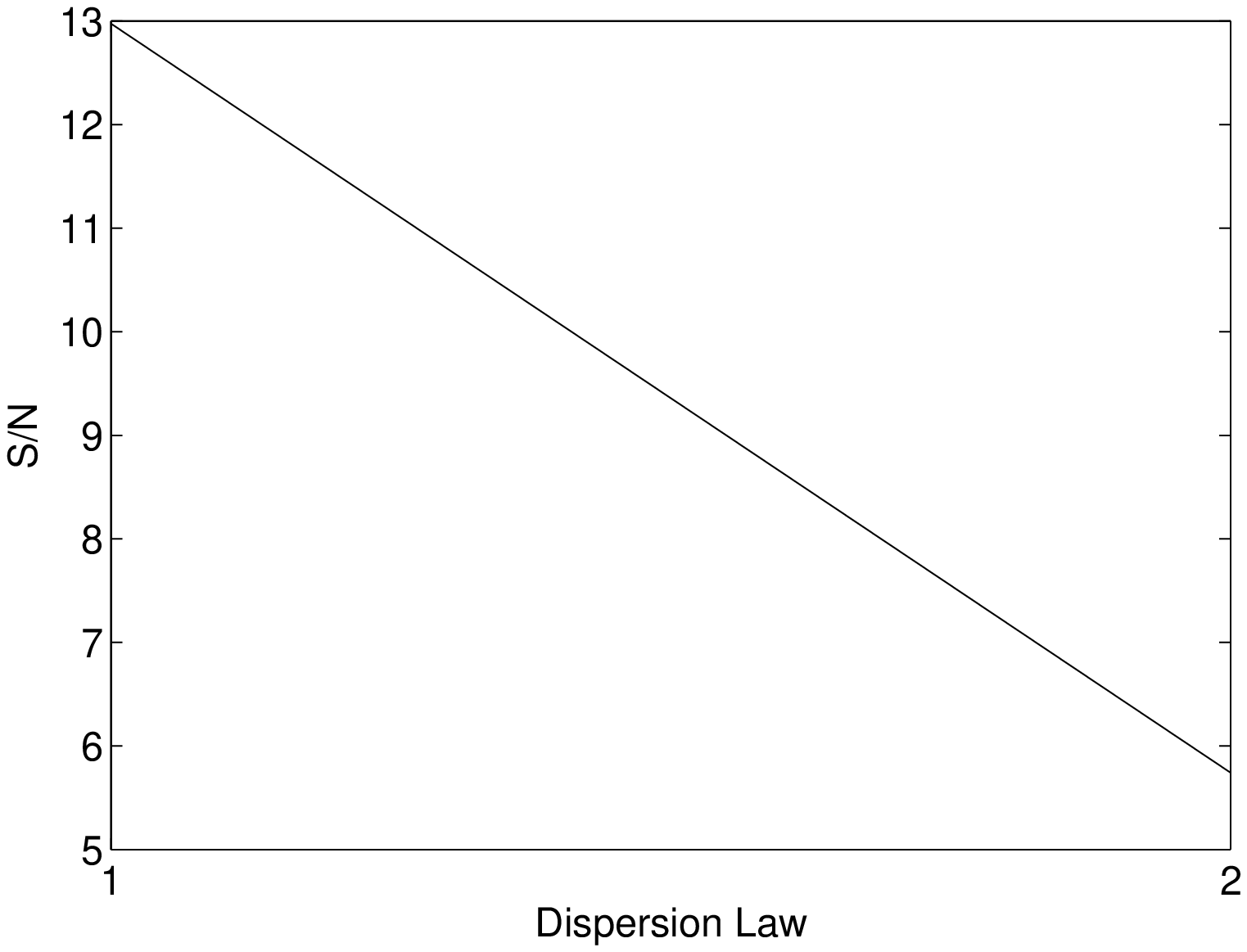,width=0.4\linewidth}}
 \caption{For one pulse of the pulsar B0833-45 observed with the Ooty Radio Telescope \mbox{((a)-(d))}, and
          for the swept-frequency RFI in data from the B1919+21 observation with the Gauribidanur Telescope \mbox{((e)-(h))},
          \mbox{(a)\&(e) maximum} S/N versus dispersion law, DM channel, and pulse width,
          \mbox{(b)\&(f) S/N} versus the DM channel corresponding to maximum S/N,
          \mbox{(c)\&(g) S/N} versus the pulse width corresponding to maximum S/N,
          \mbox{(d)\&(h) S/N} versus the dispersion law corresponding to maximum S/N.}
 \label{fig_summary}
 \end{figure}

\subsection{Narrow-band RFI}
Narrow-band RFI can be identified by analysing the smoothness of its spectrum. For continuum signals, spectral features cannot be narrower than what the order of the filter dictates. Spectral lines from astronomical sources are narrow, but they can be identified by virtue of the fact that they have the same temporal statistics as continuum signals. RFI does not follow the same statistics, leading to identification.

\subsection{Broadband RFI}
Radio astronomical signals are always dispersed, although the effect of dispersion is apparent only for suitably short-duration signals. If dispersive characteristics are absent in the signal, it can be concluded that it is of terrestrial origin. Unfortunately, the converse is not always true. Some broadband RFI mimic the dispersed nature of astronomical signals, when the RFI frequency sweeps across the observing band. Swept-frequency radars are typical sources of such RFI. The difference between this type of RFI and astronomical signals is that the former is less likely to follow the dispersion law ${\Delta t \propto f^{-2}}$ as followed by the latter, and is more likely to appear to be `dispersed' in a linear manner.

\section{Search Procedure}
Our single pulse search technique extends the original single pulse search algorithm by introducing `dispersion law' as a third parameter, in addition to time duration and DM. A `maximum S/N cube' corresponding to one candidate pulse is shown in Fig.~\ref{fig_summary}(a) and (e), with the former corresponding to the pulsar \mbox{B0833-45}, and the latter, a swept-frequency RFI in the data from an observation of the pulsar B1919+21. The maximum S/N with respect to DM channel and pulse width is represented by the mesh layers. The vertical axis marks the dispersion law. The two laws represented here are the linear (${\propto f}$, dispersion law index 1) and the inverse square law (${\propto f^{-2}}$, dispersion law index 2). The maximum of the S/N values corresponding to each law is calculated and compared, and as can be seen in Fig.~\ref{fig_summary}(d) and (h), the slope of the curve indicates whether the candidate is the result of RFI, or an astronomical transient.

\acknowledgements
We thank the Gauribidanur Telescope staff, especially H. Aswathappa, and the Ooty Radio Telescope staff for their help in the observations. Yogesh Maan acknowledges financial support from the Council for Scientific and Industrial Research.

\end{document}